\documentclass[12pt, a4paper]{article}

\usepackage{url}
\usepackage{amsmath, amssymb, amsfonts}
\usepackage{amsfonts}
\usepackage[a4paper, left=1.2in, right=1.2in, top=1.2in, bottom=1.4in]{geometry}
\usepackage[pdfstartpage=1, 
pdfauthor={Vasileios Iliopoulos and David B. Penman}, 
    unicode=false,   
    hidelinks,               
    filecolor=magenta,     
    urlcolor=blue]{hyperref}
\hypersetup{pdfstartview={XYZ null null 0.94}}

\renewcommand\footnotemark{}

\begin{document}

\begin{center}
Postprint of the article ``Dual pivot Quicksort'' published in the Journal of Discrete Mathematics, 
Algorithms and Applications, Vol. 4, No. 3 (2012) 1250041
\copyright ~World Scientific Publishing Company. \\
\href{http://www.worldscientific.com/doi/abs/10.1142/S1793830912500413}{DOI: 10.1142/S1793830912500413}.
\end{center}

\title{Dual pivot Quicksort}
\author{Vasileios Iliopoulos$^\star$ and David B. Penman$^\dagger$ \\ 
\\
Department of Mathematical Sciences \\ 
University of Essex, Wivenhoe Park \\ 
Colchester, CO4 3SQ, U.K. }
\date{}
\begingroup
\let\newpage\relax
\maketitle
\endgroup
\footnotetext{$^{\star}$E-mail: \href{mailto:viliop@essex.ac.uk}{\tt viliop@essex.ac.uk} 
}
\footnotetext{$^{\dagger}$E-mail: \href{mailto:dbpenman@essex.ac.uk}{\tt dbpenman@essex.ac.uk}
}

\begin{abstract}
\noindent
In this paper, we analyse the dual pivot Quicksort, a variant 
of the standard Quicksort algorithm, in which two pivots are 
used for the partitioning of the array. We are solving recurrences 
of the expected number of key comparisons and exchanges performed 
by the algorithm, obtaining the exact and asymptotic total 
average values contributing to its time complexity. Further, we 
compute the average number of partitioning stages and the variance 
of the number of key comparisons. In terms of mean values, dual pivot 
Quicksort does not appear to be faster than ordinary algorithm.
\end{abstract}

\section{Introduction}
Quicksort is a sorting algorithm with an extensive literature regarding its mathematical analysis and its applications. Without loss of generality, suppose that we want to quick sort a random permutation of distinct keys \{1, 2, \ldots , n\} with all the $n!$ permutations equally likely. A key is randomly chosen as pivot and by pairwise comparisons of the other elements with it, smaller keys are placed to left and greater to right. Now, the pivot $j$ is at its final position and the algorithm is recursively invoked to sort independently the left and right subarrays of $(j-1)$ and $(n-j)$ elements, respectively. Letting $C_{n}$ being the number of comparisons for sorting $n$ keys, its average value is given by the following recursive relation 
\begin{align*}
\mathbb {E} (C_{n}) & = n-1+\dfrac{1}{n}\sum_{j=1}^{n} \bigl ( \mathbb {E} (C_{j-1})+ \mathbb {E} (C_{n-j}) \bigr ) \\
& = n-1+\dfrac{2}{n}\sum_{j=1}^{n} \mathbb{E}(C_{j-1}),
\end{align*}
with initial condition $C_{0} :=  0$. Subtracting $(n-1) \mathbb {E} (C_{n-1})$ from $n\mathbb{E} (C_{n})$ and telescoping, the average number of comparisons is
\begin{align*} 
\mathbb{E}(C_{n})=2(n+1)H_{n} - 4n  \sim 2n \ln (n). \tag{1.1}
\end{align*}

Similarly, it is a routine matter to compute that the average number of
exchanges performed is 
\begin{align*}
\frac{2(n+1)H_{n}-3n}{6} \sim \dfrac{1}{3} n \ln (n),  \tag{1.2}
\end{align*}
when at the first partitioning stage, the expected number of exchanges is 
\begin{eqnarray*}
\dfrac{n}{6}+\dfrac{5}{6n}, \mbox{~\cite{2}}. 
\end{eqnarray*}
Recall that $H_{n}$ is the $n_{th}$ harmonic number, defined by
$H_{n} := \displaystyle \sum_{j=1}^{n}\frac{1}{j}$ and $H_{0} := 0$.
Further, the sign $\sim $ denotes asymptotic equivalence between two
functions $f(n)$ and $g(n)$. That is $f(n) \sim g(n)$ if and only if
$\displaystyle  \lim_{n  \to  \infty} \dfrac{f(n)}{g(n)} = 1 $. (In
\cite{2}, \cite{4}, it is suggested that small segments of size less
than some
parameter $m$ be sorted by a simpler algorithm, such as insertion sort,
as this is in practice quicker: in order to simplify the calculations for the solutions
of the recurrences, we assume that $m=0$).

\section{Partitioning on two pivots}
The idea for this variant is that we can randomly select two
elements as pivots for the partitioning of the array. The number of
comparisons obeys the following recursive rule;
\begin{eqnarray*}
C_{n}= \mbox {\sl{``Number~of~comparisons~during~first~partitioning~stage''}}+C_{i-1}+C_{j-i-1}+C_{n-j},
\end{eqnarray*}
since at the beginning, the pivots are compared each other and swapped if
they are not in order. If elements $i < j$ are selected, the array
is partitioned into three subarrays: one with $(i-1)$ keys smaller than $i$,
a subarray of $(j-i-1)$ keys between two pivots and the part of $(n-j)$ elements
greater than $j$. The algorithm then is recursively applied to each of these subarrays. The number of comparisons during the first stage is
\begin{align*}
 A_{n} = 1 + \biggl ( (i-1)+2(j-i-1)+2(n-j) \biggr ),  \quad i = 1, \ldots, n-1 \mbox{~~and~~}  j= i+1, \ldots, n.
\end{align*}
because if an element is lower than $i$, then it is less than $j$ automatically,
so $i-1$ elements beneath $i$ only need to be compared with one of the
pivots. However if an element is greater than $i$ then it needs to be
compared with the other pivot as well, to determine whether or not it is
greater than $j$. We refer to Sedgewick \cite{4} for code for a version
of this scheme. The average value of $A_{n}$ is 
\begin{align*}
& \dfrac{1}{\dbinom{n}{2}} \sum_{i=1}^{n-1}\sum_{j=i+1}^{n} \biggl ( 1 +  (i-1)+2(j-i-1)+2(n-j)  \biggr ) =
\dfrac{1}{\dbinom{n}{2}} \sum_{i=1}^{n-1}\sum_{j=i+1}^{n} 
\biggl ( 2n - i - 2 \biggr ) \\
& = \dfrac{2}{n(n-1)}\left(\dfrac{5}{6}n^3 - 2n^2+\dfrac{7}{6}n \right ) = \dfrac{5n-7}{3}.
\end{align*}

Hence, the recurrence for the expected number of comparisons is
\begin{align*}
\mathbb{E}(C_{n}) = \dfrac{5n-7}{3}+\frac{2}{n(n-1)}
\left ( \sum_{i=1}^{n-1}\sum_{j=i+1}^{n} \mathbb{E}(C_{i-1})
+\sum_{i=1}^{n-1}\sum_{j=i+1}^{n} \mathbb{E}(C_{j-i-1})
+ \sum_{i=1}^{n-1}\sum_{j=i+1}^{n} \mathbb{E}(C_{n-j}) \right ).
\end{align*}
Note that the three double sums above are equal. Therefore, the recurrence becomes
\begin{align*}
\mathbb{E}(C_{n})& = \dfrac{5n-7}{3}+\frac{6}{n(n-1)}\sum_{i=1}^{n-1}(n-i) \mathbb{E}(C_{i-1}). 
\end{align*}
Letting $a_{n} = \mathbb{E}(C_{n})$, we have 
\begin{eqnarray*}
a_{n}=\dfrac{5n-7}{3}+\frac{6}{n(n-1)}\sum_{i=1}^{n-1}(n-i)a_{i-1}, \mbox{$~n \geq 2.~$}
\end{eqnarray*}
Trivially, it holds that $a_{0} = a_{1} = 0$. Multiplying both sides by $\dbinom{n}{2}$, we obtain
\begin{align*}
\dbinom{n}{2} a_{n} & = \dbinom{n}{2} \left ( \frac{5n-7}{3} + \frac{6}{n(n-1)}\sum_{i=1}^{n-1}(n-i)a_{i-1} \right ) \\
& = \frac{n(n-1)(5n-7)}{6} + 3\sum_{i=1}^{n-1}(n-i)a_{i-1}.
\end{align*}

We introduce the difference operator $\Delta$ for the solution of this
recurrence. The operator is defined by
\begin{align*}
\Delta F(n)& :=F(n+1)-F(n) \mbox{\quad and~for~higher~orders~} \\
\Delta^{k}F(n)& :=\Delta^{k-1}F(n+1)-\Delta^{k-1}F(n).
\end{align*}
Thus, we have 
\begin{align*}
& \Delta \dbinom{n}{2} a_{n} = \dbinom{n+1}{2}a_{n+1} - \dbinom{n}{2}a_{n} = \dfrac{5n^{2}-3n}{2} + 3\sum_{i=0}^{n-1}a_{i}  \\
& \Delta^{2} \dbinom{n}{2} a_{n} = \Delta \dbinom{n+1}{2}a_{n+1} - \Delta \dbinom{n}{2}a_{n} = 5n+1+3a_{n}.
\end{align*}
By definition, 
\begin{align*}
\Delta^{2} \dbinom{n}{2} a_{n} & = \Delta \dbinom{n+1}{2}a_{n+1}- \Delta \dbinom{n}{2}a_{n} \\ 
& = \dbinom{n+2}{2}a_{n+2}-2\dbinom{n+1}{2}a_{n+1} + \dbinom{n}{2}a_{n}
\end{align*}
and the recurrence becomes
\begin{eqnarray*}
&& (n+1)(n+2)a_{n+2}-2n(n+1)a_{n+1}+n(n-1)a_{n}=2(5n+1+3a_{n}) \\
&& \Rightarrow (n+1) \bigl ((n+2)a_{n+2}-(n-2)a_{n+1} \bigr )-(n+2) \bigl ((n+1)a_{n+1}-(n-3)a_{n} \bigr )= 2(5n+1).
\end{eqnarray*}
Dividing by $(n+1)(n+2)$, we obtain the telescoping recurrence
\begin{eqnarray*}
\frac{(n+2)a_{n+2}-(n-2)a_{n+1}}{n+2}=\frac{(n+1)a_{n+1}-(n-3)a_{n}}{n+1}+\frac{2(5n+1)}{(n+1)(n+2)},
\end{eqnarray*}
which yields
\begin{eqnarray*}
\frac{(n+2)a_{n+2}-(n-2)a_{n+1}}{n+2}= 2\sum_{j=0}^{n}\frac{5j+1}{(j+1)(j+2)}=\frac{18}{n+2}+10H_{n+1}-18.
\end{eqnarray*}

The recurrence is equivalent to
\begin{eqnarray*}
na_{n}-(n-4)a_{n-1}=18+10nH_{n-1}-18n.
\end{eqnarray*}
Multiplying by $\dfrac{(n-1)(n-2)(n-3)}{24}$ , this recurrence is transformed to a telescoping one \cite{4},
\begin{eqnarray*}
\dbinom{n}{4}a_{n}=\dbinom{n-1}{4}a_{n-1}+\frac{18(n-1)(n-2)(n-3)}{24}+10 \dbinom{n}{4}H_{n-1}-18\dbinom{n}{4}.
\end{eqnarray*}
Unwinding, we have
\begin{align*}
\dbinom{n}{4}a_{n}=18 \sum_{j=1}^{n}\frac{(j-1)(j-2)(j-3)}{24}+10 \sum_{j=1}^{n}\dbinom{j}{4}H_{j-1}-18\sum_{j=1}^{n}\dbinom{j}{4}  \tag{2.1}.
\end{align*}
Using \textsc{Maple}, we found that
\begin{align*}
\sum_{j=1}^{n} (j-1)(j-2)(j-3) = 6 \dbinom{n}{4}
\end{align*}
and for the other sums in Eq. (2.1), 
\begin{eqnarray*}
&& \sum_{j=1}^{n} \dbinom{j}{4} = \dbinom{n+1}{5} \mbox{~~and~~} \sum_{j=1}^{n} \dbinom{j}{4}H_{j} = \dbinom{n+1}{5} \left (H_{n+1} - \dfrac{1}{5} \right ). 
\end{eqnarray*}
Therefore
\begin{align*}
\sum_{j=1}^{n} \dbinom{j}{4}H_{j-1} & = \sum_{j=1}^{n} \Biggl ( \dbinom{j}{4} \left ( H_{j} - \dfrac{1}{j} \right ) \Biggr ) = \sum_{j=1}^{n} \dbinom{j}{4}H_{j} - \sum_{j=1}^{n} \dbinom{j}{4} \dfrac{1}{j} \\
& = \dbinom{n+1}{5} \biggl (H_{n+1} - \dfrac{1}{5} \biggr ) - \dfrac{1}{24} \sum_{j=1}^{n}(j-1)(j-2)(j-3) \\
& = \dbinom{n+1}{5} \biggl (H_{n+1} - \dfrac{1}{5} \biggr ) - \dbinom{n}{4} \dfrac{1}{4}. 
\end{align*}

Now Eq. (2.1)  becomes
\begin{eqnarray*}
&& \dbinom{n}{4}a_{n} = \dfrac{9}{2} \dbinom{n}{4} +10 \Biggl ( \dbinom{n+1}{5} \left (H_{n+1} - \dfrac{1}{5} \right ) - \dfrac{1}{4} \dbinom{n}{4} \Biggr ) -18 \dbinom{n+1}{5}. \\
&& \Rightarrow a_{n} = \dfrac{9}{2} + 10 \Biggl ( \dfrac{n+1}{5} \left (H_{n+1} - \dfrac{1}{5} \right ) - \dfrac{1}{4} \Biggr ) - \dfrac{18(n+1)}{5}.
\end{eqnarray*}
Finally, the expected number of comparisons, when two pivots are chosen is
\begin{align*}
a_{n} = 2(n+1)H_{n}- 4n \sim 2n \ln(n).  \tag{2.2}
\end{align*}
This is exactly the same as the expected number of comparisons in Eq. (1.1) computed earlier for ordinary Quicksort. The dual pivot variant is claimed to be faster
in experimental measurements than the standard algorithm in \cite{6}.  A referee of this article commented to us that the variant gives a $30\%$ performance boost on randomly permuted data.

We proceed to compute the average number of exchanges. Letting $S_{n}$ denote the total number of exchanges we carry out
when sorting $n$ objects, we have that
$$S_{n}=\mbox{ \sl {``Number~of~exchanges~during~first~partitioning~stage''}}+
S_{i-1}+S_{j-i-1}+S_{n-j}.$$
Now it is fairly clear that, again using that the pivots are chosen uniformly
at random, that the average values of last three quantities are equal. So our main objective now is to determine the average number of exchanges during
the first partitioning stage. At the end of the partition routine, $(i-1)$ elements are less than pivot $i$. Thus, the contribution to the number of exchanges is \cite{4},
\begin{eqnarray*}
\dfrac{1}{\dbinom{n}{2}}\sum_{i=1}^{n-1}\sum_{j=i+1}^{n}(i-1)
= \dfrac{1}{\dbinom{n}{2}}\sum_{i=1}^{n-1}(n-i)(i-1)
= \dfrac{1}{\dbinom{n}{2}} \left ( \sum_{i=1}^{n-1}(n-i)i -
\sum_{i=1}^{n-1}(n-i) \right ).
\end{eqnarray*}
The first sum being evaluated gives
\begin{align*}
\sum_{i=1}^{n-1}(n-i)i =\dfrac{n^{3}-n}{6}
\end{align*}
and the second is just $n(n-1)/2$. The average contribution is 
\begin{eqnarray*}
\dfrac{2}{n(n-1)} \left ( \dfrac{n^{3}-n}{6} - \dfrac{n(n-1)}{2} \right ) = \dfrac{n-2}{3}.
\end{eqnarray*}

Similarly, the average number of exchanges for the $(n-j)$ elements greater
than the second pivot is $ \displaystyle  \left (\dfrac{n-2}{3} \right )$,
since the double sums are equal. Adding the two final ``exchanges'' to get the pivots in place, the average number of exchanges during the partitioning routine is $\left (\dfrac{2(n+1)}{3} \right )$. Therefore, the recurrence for the mean number of exchanges in course of the algorithm is
\begin{align*}
\mathbb{E}(S_{n})& =\dfrac{2(n+1)}{3}+ \frac{2}{n(n-1)} \left ( \sum_{i=1}^{n-1}\sum_{j=i+1}^{n}\mathbb{E}(S_{i-1})+\sum_{i=1}^{n-1}\sum_{j=i+1}^{n} \mathbb{E}(S_{j-i-1}) + \sum_{i=1}^{n-1}\sum_{j=i+1}^{n} \mathbb{E}(S_{n-j}) \right ) \\
& = \dfrac{2(n+1)}{3}+ \frac{6}{n(n-1)} \sum_{i=1}^{n-1}\sum_{j=i+1}^{n} \mathbb{E}(S_{i-1}) \\
& = \dfrac{2(n+1)}{3}+ \frac{6}{n(n-1)} \sum_{i=1}^{n-1}(n-i)
\mathbb{E}(S_{i-1}). 
\end{align*}
This recurrence is solved in \cite{4}: here we present a solution using
generating functions. Letting $b_{n} = \mathbb{E}(S_{n})$, we have
\begin{eqnarray*}
b_{n}=\dfrac{2(n+1)}{3}+ \frac{6}{n(n-1)} \sum_{i=1}^{n}(n-i)b_{i-1}.
\end{eqnarray*}
Multiplying by $\dbinom{n}{2}$ to clear fractions, we have
\begin{align*}
\dbinom{n}{2}b_{n} & = \dbinom{n}{2} \left (\dfrac{2(n+1)}{3}+ \frac{6}{n(n-1)} \sum_{i=1}^{n}(n-i)b_{i-1} \right ) \\
& = \dfrac{n(n-1)(n+1)}{3} + 3 \sum_{i=1}^{n}(n-i)b_{i-1}.
\end{align*}
Multiplying by $z^{n}$ and summing over $n$, in order to
obtain the generating function
$g(z)= \displaystyle \sum_{n=0}^{\infty}b_{n}z^{n}$ of the coefficients
$b_{n}$,
\begin{align*}
\sum_{n=0}^{\infty} \dbinom{n}{2}b_{n}z^{n} & = \dfrac{1}{3} \sum_{n=0}^{\infty}n(n-1)(n+1)z^{n}+3 \sum_{n=1}^{\infty} \sum_{i=1}^{n}(n-i)b_{i-1}z^{n} \\
\Rightarrow \dfrac{z^{2}}{2} \sum_{n=0}^{\infty}n(n-1)b_{n}z^{n-2} & = \dfrac{z^{2}}{3} \sum_{n=0}^{\infty}n(n-1)(n+1)z^{n-2}+3 \sum_{n=1}^{\infty} \sum_{i=1}^{n}(n-i)b_{i-1}z^{n} \\
\Rightarrow \dfrac{z^{2}}{2} \dfrac{\operatorname{d}^{2} g(z)}{\operatorname{d}z^{2}} & =\dfrac{z^{2}}{3} \dfrac{\operatorname{d}^{3}}{\operatorname{d}z^{3}} \left (\displaystyle \sum_{n=0}^{\infty}z^{n+1} \right ) +3 \sum_{n=1}^{\infty}\sum_{i=1}^{n}(n-i)b_{i-1}z^{n}.
\end{align*}

The first term on the right-hand side of the previous equation is the third order derivative of the following geometric series
\begin{eqnarray*}
\sum_{n=0}^{\infty}z^{n+1}=\dfrac{z}{1-z}
\mbox{~~~and~~~} \dfrac{\operatorname{d}^{3}}{\operatorname{d}z^{3}} \left ( \dfrac{z}{1-z} \right ) = \dfrac{6}{(1-z)^{4}}, \mbox{~$\vert z \vert < 1$}.
\end{eqnarray*}
The double sum is equal to
\begin{align*}
\sum_{n=1}^{\infty}\sum_{i=1}^{n}(n-i)b_{i-1}z^{n} & = b_{0}z^{2} + ( 2b_{0}+b_{1} )z^{3} + ( 3b_{0} + 2b_{1} +b_{2} )z^{4} + \ldots \\
& = z^{2}(b_{0}+b_{1}z+b_{2}z^{2}+ \ldots ) + 2z^{3}(b_{0}+b_{1}z+b_{2}z^{2}+ \ldots ) + \ldots \\
& = (z^{2} + 2z^{3} + 3z^{4} + \ldots )g(z) \\
& = \biggl ( \sum_{n=0}^{\infty} nz^{n+1} \biggr )g(z).
\end{align*}
The sum which multiplies $g(z)$ on the last line is  
\begin{eqnarray*}
\sum_{n=0}^{\infty} nz^{n+1}=\left ( \dfrac{z}{1-z} \right )^{2}.
\end{eqnarray*}
Therefore, our recurrence is transformed to the following differential equation
\begin{eqnarray*}
\dfrac{z^{2}}{2}\dfrac{\operatorname{d}^{2} g(z)}{\operatorname{d}z^{2}}=\dfrac{2z^{2}}{(1-z)^{4}} + 3 g(z) \left ( \dfrac{z}{1-z} \right )^{2}.
\end{eqnarray*}
Changing variables $v=1-z$, we have that $f(v)= g(1-v)$. Thus, it holds
\begin{eqnarray*}
\dfrac{\operatorname{d}^{k}f(v)}{\operatorname{d}v^{k}}= (-1)^{k}\dfrac{\operatorname{d}^{k}g(1-v)}{\operatorname{d}v^{k}}.
\end{eqnarray*}
The differential equation becomes
\begin{eqnarray*}
\dfrac{(1-v)^{2}}{2}\dfrac{\operatorname{d}^{2} f(v)}{\operatorname{d}v^{2}}=\dfrac{2(1-v)^{2}}{v^{4}} + 3 f(v) \left ( \dfrac{1-v}{v} \right )^{2}.
\end{eqnarray*}

Using \textsc{Maple}, the general solution is
\begin{eqnarray*}
f(v)=c_{1}v^{3}+ \dfrac{c_{2}}{v^{2}} - \dfrac{20 \ln (v)+4}{25v^{2}}, \mbox{~$c_{1}, ~ c_{2} \in \mathbb R$.}
\end{eqnarray*}
For the computation of constants, we consider the fact that $f(1) = g(0) = 0$ \mbox{~
and ~} $f^\prime(1)=- g^\prime(0) = 0$ ~ (as $b_{0}=b_{1}=0$). The resulting
system of linear equations in $c_{1}$ and $c_{2}$ has solution
$(c_{1}, c_{2} ) =  \left ( \dfrac{4}{25}, ~ 0 \right )$. Therefore, the function is
\begin{eqnarray*}
f(v)= \dfrac{4}{25}v^{3} - \dfrac{20 \ln (v)+4}{25v^{2}}.
\end{eqnarray*}
But, since 
\begin{eqnarray*}
(1-z)^{-2} = \sum_{n=0}^{\infty}(n+1)z^{n}\mbox{~~and~~}
\dfrac{\ln (v)}{v^{2}}=
  - \dfrac{1}{(1-z)^{2}} \ln \left (\dfrac{1}{1-z} \right ),
\end{eqnarray*}
this can be written as a product of the following two series;
\begin{align*}
\dfrac{1}{(1-z)^{2}} \ln \left ( \dfrac{1}{1-z} \right ) & = \left (\sum_{n=1}^{\infty}nz^{n-1} \right ) \left (\sum_{n=1}^{\infty}\dfrac{1}{n}z^{n} \right ) \\
& = \biggl ( 1+2z+3z^{2}+4z^{3}+ \ldots \biggr )\left(  z+\dfrac{z^{2}}{2}+\dfrac{z^{3}}{3}+\dfrac{z^{4}}{4} + \ldots \right ) \\
& = z + \biggl ( 2 + \dfrac{1}{2} \biggr )z^{2} + \biggl ( 3 + \dfrac{2}{2} + \dfrac{1}{3} \biggr )z^{3} + \biggl ( 4 + \dfrac{3}{2} + \dfrac{2}{3} + \dfrac{1}{4} \biggr )z^{4} + \ldots  \\
& =  z + \bigl (H_{1}+H_{2} \bigr )z^{2} + \bigl (H_{1}+H_{2}+H_{3} \bigr )z^{3} + \ldots  \\
& = \sum_{n=0}^{\infty} \biggl ( (n+1)H_{n}-n \biggr )z^{n}.
\end{align*}
Extracting the coefficients and discarding terms for $n \leq 3$, the exact mean number of exchanges of dual pivot Quicksort is equal to, 
\begin{align*}
b_{n} & = \dfrac{4}{5} \biggl ( (n+1)H_{n}-n \biggr ) - \dfrac{4}{25} (n+1) \\ 
& = \dfrac{4}{5}(n+1)H_{n} - \dfrac{24n+4}{25} \sim \dfrac{4}{5} n \ln (n).  \tag{2.3} 
\end{align*}
Comparing the expected number of comparisons of this variant with the standard algorithm, we see that they are identical. However, the mean number of exchanges is $2.4$ times greater than the figure of normal Quicksort. 

In the lines that follow, we compute the average number of partitioning stages $\mathbb{E}(P_{n})$ of dual pivot Quicksort. The recurrence is much
simpler;
\begin{eqnarray*}
P_{n} = 1  + P_{i-1} + P_{j-i-1}+P_{n-j}.
\end{eqnarray*}
Averaging over all $\dbinom{n}{2}$ pairs of pivots $i$ and $j$ 
we have
\begin{align*}
\mathbb{E}(P_{n})& =  1  + \dfrac{6}{n(n-1)}\sum_{i=1}^{n-1}\sum_{j=i+1}^{n} \mathbb{E}(P_{i-1}) \\
& = 1 + \dfrac{6}{n(n-1)}\sum_{i=1}^{n-1}(n-i) \mathbb{E}(P_{i-1}),
\end{align*}
since the sums are equal.
Again, we use generating functions for the solution of the recurrence. Letting $f(z) = \displaystyle \sum_{n=0}^{\infty} \mathbb{E}(P_{n})z^{n}$ our recurrence is transformed to the following differential equation
\begin{eqnarray*}
f''(z)\dfrac{z^{2}}{2} = \dfrac{z^{2}}{(1-z)^3} + 3f(z) \left (\dfrac{z}{1-z}  \right )^{2}.
\end{eqnarray*}  
Changing variables $x = 1-z$ we have $h(x) = f(1-x)$ and the general solution is $$h(x) = \dfrac{c_{2}}{x^{2}}+x^{3}c_{1}-\dfrac{1}{2x}.$$
Since $h(1)=h^{\prime}(1) = 0$ the constants are $c_{1} = \dfrac{1}{10}$ and $c_{2} = \dfrac{2}{5}$. 
Consequently, the mean number of partitioning stages is found to be equal to 
\begin{align}
\mathbb{E}(P_{n}) = \dfrac{2}{5}(n+1)- \dfrac{1}{2}.    \tag{2.4}
\end{align} 
This is smaller than the expected number of stages of ordinary Quicksort, which is $n$, when there is no switch to straight insertion for the sorting of small subfiles, \cite{4}. 

\section{Variance}
Finally, we set up a recurrence for the computation of the variance of the number of comparisons of dual pivot Quicksort. Recall that
\begin{eqnarray*}
A_{n} = 2n- i - 2 \mbox{~~~and~~~} 
\operatorname{\mathbb{E}}(A_{n}) = \dfrac{5n-7}{3}.
\end{eqnarray*}
By the recurrence relation for the number of comparisons, we have
\begin{eqnarray*}
P(C_{n} = t) = \dfrac{1}{\dbinom{n}{2}}\sum_{i=1}^{n-1}\sum_{j=i+1}^{n}P(C_{n} = t) = \dfrac{1}{\dbinom{n}{2}}\sum_{i=1}^{n-1}\sum_{j=i+1}^{n}P(A_{n}+C_{i-1}+C_{j-i-1}+C_{n-j} = t),
\end{eqnarray*}
noting that the resulting subarrays are independently sorted, the above is
\begin{eqnarray*}
\dfrac{1}{\dbinom{n}{2}}\sum_{i=1}^{n-1}\sum_{j=i+1}^{n}\sum_{l, m}\biggl (P(C_{i-1} = l)  P(C_{j-i-1} = m)P(C_{n-j} =  t-m-l-2n+i+2) \biggr).
\end{eqnarray*}
Letting $\displaystyle f_{n}(z) = \sum_{t = 0}^{\infty}P(C_{n} = t)z^{t}$ be the ordinary probability generating function for the number of comparisons needed to sort $n$ keys, we obtain
\begin{align*}
f_{n}(z) = \dfrac{1}{\dbinom{n}{2}}\sum_{i=1}^{n-1}\sum_{j=i+1}^{n}z^{2n-i-2}f_{i-1}(z)f_{j-i-1}(z)f_{n-j}(z).  \tag{3.1}
\end{align*}
It holds that $f_{n}(1) = 1$ and $f'_{n}(1) = 2(n+1)H_{n} - 4n$. Moreover, the second order derivative of Eq. (3.1) evaluated at $z=1$ is recursively given by
\begin{eqnarray*}
&& f''_{n}(1)~ = \dfrac{2}{n(n-1)} \biggl ( \sum_{i=1}^{n-1}\sum_{j=i+1}^{n}(2n-i-2)^{2}- \sum_{i=1}^{n-1}\sum_{j=i+1}^{n}(2n-i-2) \\
&& +~2\sum_{i=1}^{n-1}\sum_{j=i+1}^{n}(2n-i-2)E(C_{i-1}) + 2\sum_{i=1}^{n-1}\sum_{j=i+1}^{n}(2n-i-2)E(C_{j-i-1}) \\
&& +~2\sum_{i=1}^{n-1}\sum_{j=i+1}^{n}(2n-i-2)E(C_{n-j})+2\sum_{i=1}^{n-1}\sum_{j=i+1}^{n}E(C_{i-1})E(C_{j-i-1}) \\
&& + ~2\sum_{i=1}^{n-1}\sum_{j=i+1}^{n}E(C_{i-1})E(C_{n-j}) +2\sum_{i=1}^{n-1}\sum_{j=i+1}^{n}E(C_{j-i-1})E(C_{n-j}) \\
&& +~\sum_{i=1}^{n-1}\sum_{j=i+1}^{n}f''_{i-1}(1) +  \sum_{i=1}^{n-1}\sum_{j=i+1}^{n}f''_{j-i-1}(1)+\sum_{i=1}^{n-1}\sum_{j=i+1}^{n}f''_{n-j}(1) \biggr ).
\end{eqnarray*}

The reader should not be discouraged by this long expression, since many of the sums are equal. Specifically, the fourth and fifth turn out to be equal and by simple manipulation of indices, the sums of the products of expected values are equal. The double sum of the product of the mean number of comparisons can be simplified as follows:
\begin{align*}
& \sum_{i=1}^{n-1}\sum_{j=i+1}^{n}E(C_{i-1})E(C_{n-j}) = \sum_{i=1}^{n-1}\left( E(C_{i-1}) \biggl (\sum_{j=0}^{n-i-1}E(C_{j}) \biggr) \right) \\
& = \sum_{i=1}^{n-1} \Biggl (\biggl( (2iH_{i-1}-4(i-1) \biggr ) 
\biggl (2 \dbinom{n-i+1}{2}H_{n-i}+ \dfrac{n-i-5(n-i)^{2}}{2} \biggr) \Biggr ).
\end{align*}
The next sum was computed using results from a paper \cite{5}, which contains interesting identities and properties of sums involving harmonic numbers.
\begin{align*}
\sum_{i=1}^{n-1}i\dbinom{n-i+1}{2}H_{i-1}H_{n-i} &= \sum_{i=1}^{n-1}\biggl((i-1) + 1 \biggr)\dbinom{n-i+1}{2}H_{i-1}H_{n-i} \\
 &= \sum_{i=1}^{n-1}(i-1) \dbinom{n-i}{2}H_{i-1}H_{n-i} +\sum_{i=1}^{n-1}\dbinom{n-i}{2}H_{i-1}H_{n-i} \\
+ & \sum_{i=1}^{n-1}(i-1)(n-i)H_{i-1}H_{n-i} + \sum_{i=1}^{n-1}(n-i)H_{i-1}H_{n-i}.
\end{align*}
The four sums can be evaluated using Corollary $3$ in \cite{5}.

After some computations in our \textsc{Maple} worksheet\footnote{The \textsc{Maple} worksheet containing the computations for the variance can be found at the web page: \url{http://www.essex.ac.uk/maths/staff/profile.aspx?ID=1326}}, the recurrence is 
\begin{align*}
f''_{n}(1)& = 2(n+1)(n+2)(H^{2}_{n} - H^{(2)}_{n}) - H_{n} \left (\dfrac{17}{3}n^{2} + \dfrac{47}{3}n + 6 \right ) + \dfrac{209}{36}n^{2} + \dfrac{731}{36}n + \dfrac{13}{6} \\
+ & \dfrac{6}{n(n-1)}\sum_{i=1}^{n-1}(n-i)f''_{i-1}(1),
\end{align*}
where $H^{(2)}_{n}$ is the second order harmonic number defined by $H^{(2)}_{n}:= \displaystyle \sum_{k=1}^{n}\dfrac{1}{k^{2}}.$ Letting $d_{n} = f''_{n}(1)$ and subtracting $\dbinom{n}{2}d_{n}$ from $\dbinom{n+1}{2}d_{n+1}$, we have
\begin{align*}
\Delta \dbinom{n}{2}d_{n}& = 4n(n+1)(n+2)(H^{2}_{n} - H^{(2)}_{n}) - \dfrac{nH_{n}}{9}(84n^{2} + 198n + 42)+ 3\displaystyle \sum_{i=1}^{n}d_{i-1} \\
+ & \dfrac{n}{9}(79n^{2} + 231n + 14 ),
\end{align*}
using the identity \cite{4}
\begin{eqnarray*}
H^{2}_{n+1} - H^{(2)}_{n+1} = H^{2}_{n} - H^{(2)}_{n} + \dfrac{2H_{n}}{n+1}.
\end{eqnarray*}

Further, it holds that 
\begin{eqnarray*}
\Delta^{2} \dbinom{n}{2}d_{n} = 12(n+1)(n+2)(H^{2}_{n} - H^{(2)}_{n}) -H_{n}(20n^{2} + 32n - 12) + 17n^{2} +37n + 3d_{n}.
\end{eqnarray*}
The previous equation is the same as
\begin{eqnarray*}
\dbinom{n+2}{2}d_{n+2}-2\dbinom{n+1}{2}d_{n+1} + \dbinom{n}{2}d_{n}
\end{eqnarray*}
and our recurrence becomes
\begin{align*}
& (n+1)(n+2)d_{n+2}-2n(n+1)d_{n+1}+n(n-1)d_{n} \\
& = 2 \Biggl ( 12(n+1)(n+2)(H^{2}_{n} - H^{(2)}_{n}) - H_{n}(20n^{2} + 32n - 12) + 17n^{2} +37n + 3d_{n} \Biggr).
\end{align*}
Dividing by $(n+1)(n+2)$, we obtain the telescoping recurrence

\begin{align*}
\frac{(n+2)d_{n+2}-(n-2)d_{n+1}}{n+2}&= \frac{(n+1)d_{n+1}-(n-3)d_{n}}{n+1} \\
+ & 2 \Biggl( 12(H^{2}_{n} - H^{(2)}_{n}) -\dfrac{ H_{n}(20n^{2} + 32n - 12)}{(n+1)(n+2)} + \dfrac{17n^{2} +37n}{(n+1)(n+2)}  \Biggr )
\end{align*}
with solution
\begin{align*}
(n+2)d_{n+2}-(n-2)d_{n+1} &= (24n^{2} +100n + 104)(H^{2}_{n+1}-H^{(2)}_{n+1}) \\
- &  H_{n+1}(88n^{2} + 292n + 224) + 122n^{2} +  346n + 224, 
\end{align*}
which is equivalent to
\begin{align*}
nd_{n} - (n-4)d_{n-1} &= (24n^{2} + 4n)(H^{2}_{n-1}-H^{(2)}_{n-1}) \\
- & H_{n-1}(88n^{2}- 60n - 8)+ 122n^{2} - 142n + 20.
\end{align*}

Again as before, multiplying both sides by $\dfrac{(n-1)(n-2)(n-3)}{24}$, the recurrence telescopes with solution 
\begin{eqnarray*}
f''_{n}(1) = 4(n+1)^{2}(H^{2}_{n+1} - H^{(2)}_{n+1}) - 4 H_{n+1}(n+1)(4n+3) + 23n^{2} + 33n + 12.
\end{eqnarray*}
Using the well known fact that $$\operatorname{ Var}(C_{n}) = f''_{n}(1) + f'_{n}(1) - \bigl (f'_{n}(1) \bigr )^{2},$$  the variance of the number of key comparisons of dual pivot Quicksort is
\begin{align*}
7n^{2}- 4(n+1)^{2} H^{(2)}_{n} - 2(n+1)H_{n} + 13n.  \tag{3.2}  
\end{align*}
The asymptotic figure is
\begin{align*}
\left( 7 - \dfrac{2}{3}\pi^{2} \right )n^{2} - 2n\ln (n) + O (n).  \tag{3.3}  
\end{align*}

Note that the variance of dual pivot Quicksort is identical with the variance of the ordinary algorithm - see Eq. (32) in \cite{3}. 
Also, in this paper we showed that the dual pivot Quicksort variant has
the same expected number of key comparisons as the standard algorithm
and as one might expect, the mean number of stages is smaller than the respective figure of the ordinary algorithm. 
However, the expected number of exchanges is notably large. An efficient partitioning procedure is described in a paper 
written by Frazer and McKellar \cite{1}, where they present and analyse the Samplesort algorithm. It is shown \cite{1} that the expected number of comparisons of Samplesort 
slowly approaches the Information - theoretic lower bound.

\end{document}